\documentclass[runningheads]{llncs}
\usepackage{graphicx}
\usepackage{array}
\usepackage{subfigure}
\usepackage{setspace}
\usepackage{amssymb}
\usepackage{ulem}
\usepackage{booktabs}
\usepackage[misc,geometry]{ifsym} 
\usepackage{amsmath}
\usepackage{multirow}
\usepackage{verbatim}

\usepackage[colorlinks,
            linkcolor=green,
            anchorcolor=red,
            citecolor=blue
            ]{hyperref}

\begin{document}

\title{TR-GAN: Topology Ranking GAN with Triplet Loss for Retinal Artery/Vein Classification}

\titlerunning{TR-GAN: Topology Ranking GAN with Triplet Loss}

\author{Wenting Chen \inst{1,2}  \and
Shuang Yu \inst{1}\textsuperscript{(\Letter)}  \and 
Junde Wu \inst{1}  \and
Kai Ma \inst{1} \and
Cheng Bian \inst{1}  \and
Chunyan Chu \inst{1}  \and
Linlin Shen \inst{2}  \and
Yefeng Zheng \inst{1}
}

\authorrunning{W. Chen et al.}
%

\institute{	
Tencent Healthcare, Tencent, Shenzhen, China \\
\email{shirlyyu@tencent.com} \\
\and
 School of Computer Science \& Software Engineering, Shenzhen University, Shenzhen, China\\
}
\maketitle              
%
\begin{abstract}

Retinal artery/vein (A/V) classification lays the foundation for the quantitative analysis of retinal vessels, which is associated with potential risks of various cardiovascular and cerebral diseases. The topological connection relationship, which has been proved effective in improving the A/V classification performance for the conventional graph based method, has not been exploited by the deep learning based method. In this paper, we propose a Topology Ranking Generative Adversarial Network (TR-GAN) to improve the topology connectivity of the segmented arteries and veins, and further to boost the A/V classification performance. A topology ranking discriminator based on ordinal regression is proposed to rank the topological connectivity level of the ground-truth, the generated A/V mask and the intentionally shuffled mask. The ranking loss is further back-propagated to the generator to generate better connected A/V masks. In addition, a topology preserving module with triplet loss is also proposed to extract the high-level topological features and further to narrow the feature distance between the predicted A/V mask and the ground-truth. The proposed framework effectively increases the topological connectivity of the predicted A/V masks and achieves state-of-the-art A/V classification performance on the publicly available AV-DRIVE dataset.


\end{abstract}
%
%
\section{Introduction}

Morphological changes of retinal arteries and veins have been reported to be associated with potential risks of various systemic, cardiovascular and cerebral diseases  \cite{wong2002retinal,chew2012retinal}. For example, it is reported that narrowing of arteries and widening of veins are associated with the increased risk of stroke \cite{ikram2004retinal}. 
In addition, research has also found that the narrowing of retinal arteriolar caliber is related to the risk of hypertension and developing diabetes \cite{chew2012retinal,nguyen2008retinal}. 
Therefore, accurate A/V classification is of strong clinical interest.

Automatic A/V classification has been actively investigated in recent years.
Previous works are generally performed in a two-stage manner \cite{Niemeijer2010,xu_improved_2017,dashtbozorg_automatic_2014,estrada_retinal_2015,zhao_retinal_2018,zhao2019retinal}, which requires vessel segmentation in the first stage and then classifies individual pixels on the vessel or centerline into either A/V. Depending on whether topological structure is used in the procedure, those works can be further categorized into pure feature based methods \cite{Niemeijer2010,xu_improved_2017} and graph based methods \cite{dashtbozorg_automatic_2014,estrada_retinal_2015,zhao_retinal_2018,zhao2019retinal}. The graph based methods first reconstruct the graph tree structures for individual vessels, and then take advantage of the topological connection/crossover relationship to further refine the A/V classification results, which have been proved to be effective in boosting A/V classification performance \cite{dashtbozorg_automatic_2014,estrada_retinal_2015,zhao_retinal_2018,zhao2019retinal}.

Very recently, deep learning has been introduced to the A/V classification task, which enables the end-to-end A/V classification and eliminates the pre-requirement of vessel segmentation in the first place. 
For example, Xu \textit{et al.} \cite{xu2018simultaneous} first proposed to adopt the U-Net framework to segment arteries and veins from the background.  AlBadawi and Fraz \cite{albadawi_arterioles_2018} performed pixel-wise A/V classification with an encoder-decoder based fully convolutional network. Ma \textit{et al.} \cite{ma2019multi} proposed to perform retinal vessel segmentation and A/V classification simultaneously with a multi-task framework, which achieved the state-of-the-art performance on the publicly available AV-DRIVE dataset. 

Deep learning based methods, although achieved high A/V classification accuracy, still suffer from the problem of limited connectivity for the obtained arteries and veins, which should be connected naturally. As aforementioned, the topological connection relationship of individual vessels can be adopted to mitigate this issue. To the best of our knowledge, however, integrating the vessel topological connectivity into the deep learning based A/V classification framework was rarely investigated. There is still a research gap of how to increase the topological connectivity of the predicted arteries and veins.

In order to fill this gap, a Topology Ranking Generative Adversarial Network (TR-GAN) is proposed in this paper, which tries to increase the topological connectivity of arteries and veins, and further improves the A/V classification performance. This paper is the first work trying to increase topology connectivity for the A/V classification task with a deep learning based framework.
Three major contributions are made with this paper. Firstly, instead of using a general discriminator in adversarial learning that distinguishes the ground-truth (real) mask from the generated (fake) mask, a topology ranking discriminator based on ordinal regression is proposed to rank the topological connectivity of the real mask, the fake mask and the intentionally shuffled mask. The ranking loss is further back-propagated to encourage the segmentation network to produce A/V masks with higher topological connectivity. Secondly, a topology preserving module with triplet loss is also proposed to extract the high-level topological features from the generated mask, based on which further to narrow the feature distance between the generated mask and the ground-truth mask. 
Last but not least, the proposed framework achieves the state-of-the-art performance on the publicly available AV-DRIVE dataset for the A/V classification, surpassing the most recent approaches with a remarkable margin.


\section{Method}

\begin{figure}[t]
\centering
\includegraphics[width=11.0cm,height=6.2cm]{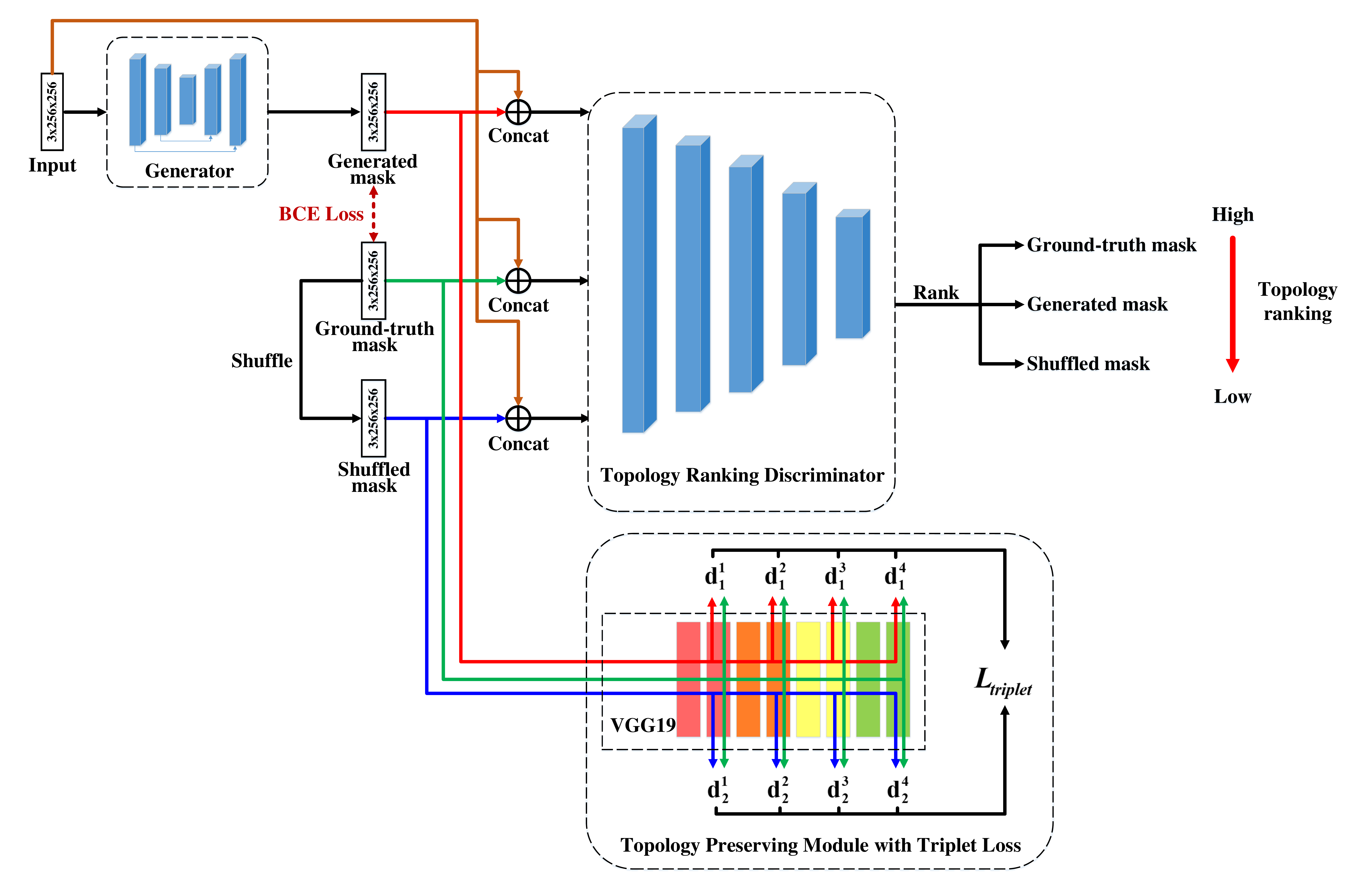}
\caption{Architecture of the proposed framework, including a generator based on U-Net, a topology ranking discriminator and a topology preserving module with triplet loss.} 
\label{fig_Framework}
\end{figure}

Fig. \ref{fig_Framework} shows the architecture of the proposed Topology Ranking GAN (TR-GAN) framework for the retinal A/V classification task.
The overall architecture consists of three parts: (1) the segmentation network as the generator, (2) the topology ranking discriminator and (3) the topology preserving module with triplet loss. 
For the segmentation network, we adopt the widely used U-Net \cite{Ronneberger2015UNetCN} architecture with a pretrained ResNet-18 as the encoder, and the decoder outputs a three-channel probability map for artery, vein and vessel segmentation. 
The generated mask is then concatenated with the original input image and fed to the topology ranking discriminator to evaluate the topological connectivity, similar for the ground-truth mask and the shuffled mask. A topology ranking adversarial loss is then calculated and back-propagated to update the generator. 

Moreover, these three masks are also fed to the topology preserving module, which extracts the high-level topological features with a pretrained VGG-19 backbone \cite{Simonyan2014VeryDC}. The extracted features are further used to construct a triplet loss that tries to narrow the feature distance between the ground-truth mask and the generated mask.

\subsection{Topology Ranking Discriminator}

In this paper, the discriminator adopts the PatchGAN structure \cite{isola2017image} and contains six convolutional layers. Input to the discriminator is conditioned by concatenating the input images and the A/V masks, so as to ensure that the generated masks are relevant to the corresponding original image content.

\begin{figure}[t]
	\centering
	\includegraphics[width=11.5cm]{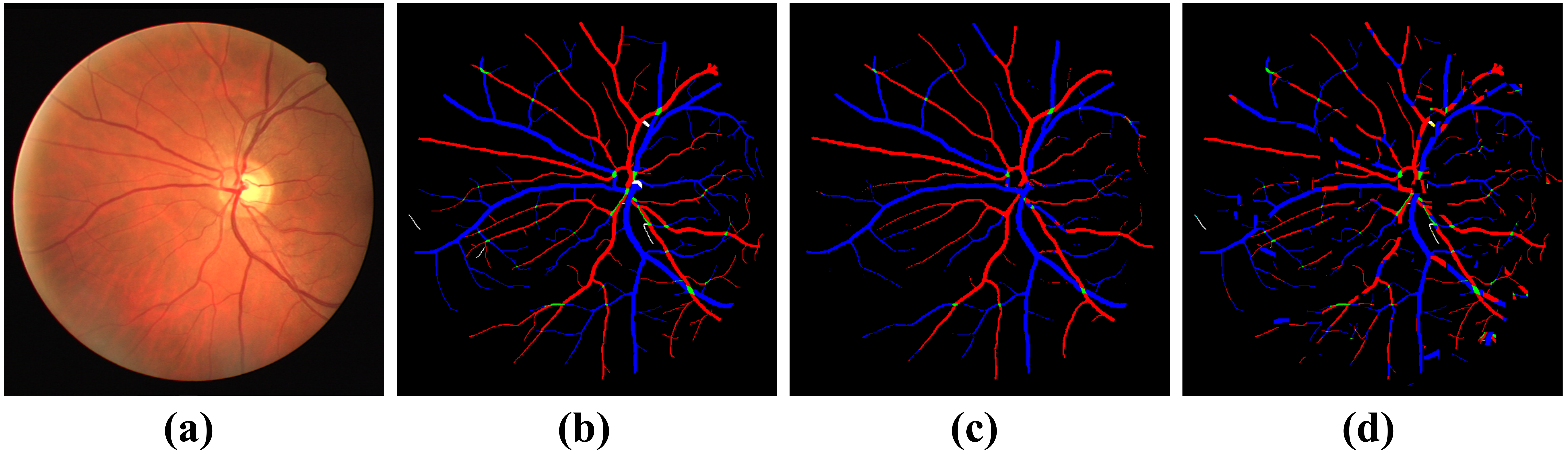}
	\caption{Different levels of topology connectivity for artery/vein classification. Red denotes arteries, blue for veins and green for the cross-overs. (a) Original color fundus image; (b) ground-truth mask with highest level of topology connectivity; (c) model prediction with median level of topology connectivity; (d) shuffled ground-truth mask with lowest level of topology connectivity.} 
	\label{fig_topoRank}
\end{figure}

Instead of using a typical discriminator which distinguishes the real masks (ground-truth) from the fake ones (the segmented masks), we propose a topology ranking discriminator to rank the level of topology connectivity for the masks. In order to enrich potential connectivity levels and synthesize a mask with low topology connectivity, we intentionally shuffle the ground-truth mask by removing part of the vessels, shifting the position of vessel segments or exchanging the A/V labels for randomly selected regions. This random shuffle procedure is repeated until 5-25$\%$ of the vessel pixels shuffled away from the ground-truth vessels, so as to ensure that the shuffled mask has a lower level of topology connectivity compared with the generated mask. Afterwards, we have three different levels of topology connectivity ranked in descending order as: the ground-truth mask, the generated mask and the shuffled mask, as demonstrated in Fig. \ref{fig_topoRank}.

In order to preserve the ranking information contained in the three different classes, ordinal classification \cite{frank2001simple} is adopted to rank the three categories. For a simple yet effective implementation of the ordinal classification, the ordinal labels can be transferred to a multi-label classification problem with two individual labels: shuffled mask $ y_1: [0, 0]$, generated mask $y_2:  [1, 0]$ and ground-truth mask $ y_3: [1, 1]$. The first element indicates whether the mask topology is better than the shuffled mask, and the second denotes whether the mask topology is better than the generated mask. The discriminator can then be optimized with:

\begin{equation}
\label{equ1}
\begin{aligned}
\mathcal{L}^D_{adv} = & \mathbb{E}_{L_s, y_1} [- y_1 \cdot log D(x,L_s ) ] + \\ 
& \mathbb{E}_{G(x), y_2} [- y_2 \cdot log D(x,G(x)) ] + \\ 
& \mathbb{E}_{L, y_3} [- y_3 \cdot log D(x,L ) ],
\end{aligned}
\end{equation}
where $x$ denotes the input image, and $L_s, G(x), L$ represent the shuffled mask, generated mask and ground-truth mask, respectively. By minimizing this objective, the discriminator $D$ learns to rank the topology connectivity levels.

When updating the generator, the optimization target is to encourage the generated mask to have the highest level of topology and fool the discriminator, with the adversarial loss as below:
\begin{equation}
\label{equ2}
\mathcal{L}^G_{adv} = \mathbb{E}_{x, y_3} [- y_3 \cdot log D(x,G(x) ) ] .
\end{equation}

\subsection{Topology Preserving Module with Triplet Loss}

Pixel-wise losses, including the binary cross entropy loss and L1/L2 loss, can only measure the low-level pixel-wise difference between the segmented mask and the ground-truth,  which often fail to measure the high-level topology difference. Therefore, we propose to calculate the distance of features extracted by a pretrained model, which is expected to contain high-level topological information \cite{mosinska2018beyond}. In this paper, a VGG-19 backbone pretrained on ImageNet is adopted to extract the feature maps. A triplet loss is further proposed to narrow the feature distance between the ground-truth mask and the generated mask.

For the triplet loss configuration, the ground-truth mask $L$ is selected as the anchor exemplar, the generated mask $G(x)$ as the positive exemplar and the shuffled mask $L_s$ as the negative exemplar. 
The triplet loss is defined as:
\begin{equation}
\label{equ8}
L_{triplet} = \frac{1}{N}\sum_{i=1}^{N} max \left ( d_{1}^i - d_{2}^i + \alpha, 0 \right ),
\end{equation}
where $\alpha$ is the margin hyper-parameter and set to 1 by default in this research; $N$ denotes the total number of network layers where we extract feature maps from. In this paper $N$ is 4 and feature maps are extracted from the ReLU layers of the $2^{nd}, 4^{th}, 6^{th}$ and $8^{th}$ convolutional blocks.
$d_1^i$ and $d_2^i$ represent the distance of anchor-positive and anchor-negative pairs for the $i^{th}$ layer of the feature extraction network, and can be obtained with:

\begin{equation}
\label{equ4}
d_1^i= \frac{1}{C_i W_i H_i}||\mathcal{F}_i (L)-\mathcal{F}_i (G(x)) ||_2^2,
\end{equation}
\begin{equation}
\label{equ5}
d_2^i= \frac{1}{C_i W_i H_i}||\mathcal{F}_i (L)-\mathcal{F}_i (L_s) ||_2^2,
\end{equation}
where $C_i,W_i,H_i$ denote the channel number, width and height, respectively, for the feature maps of the $i^{th}$ layer; $\mathcal{F}_i$ denotes the mapping function for the $i^{th}$ model layer.

\subsection{Generator Loss Function}

In order to optimize the generator, the widely adopted cross entropy loss is utilized to optimize the pixel-level difference between the generated mask and the ground-truth mask.
Meanwhile, the adversarial loss and triplet loss are also used to optimize the high-level topological information for the generated mask. The final loss for the generator network is the weighted combination of all three losses:
\begin{equation}
\label{equ6}
L_G= L_{BCE} + \lambda_1 L_{adv}^G + \lambda_2 L_{triplet},
\end{equation}
\begin{equation}
\label{equ7}
L_{BCE}=-\sum_{c=1}^{3} \mu_{c}  L_{c} \log (G(x)),
\end{equation}
where $c$ denotes the $c^{th}$ class of the output; the weights of vessel, artery and vein are denoted as $\mu_{c}$ and empirically set as 0.4, 0.3 and 0.3, respectively; $\lambda_1$ and $\lambda_2$ are the hyper-parameters to control the relative importance of each loss compared to the binary cross entropy loss and empirically set as 0.2 and 0.1, respectively.


\section{Experimental Results}

The proposed model is trained and tested on the publicly available datasets, including the AV-DRIVE dataset \cite{Hu2013_drive} and INSPIRE-AVR dataset \cite{INSPIRE2011}. The AV-DRIVE dataset contains 20 training images and 20 test images with the dimension of $584\times565$ pixels and a field-of-view of 45 degrees. Pixel-wise annotation of artery and vein is provided by AV-DRIVE. Besides, we have also tested the model on the INSPIRE-AVR dataset, which contains 40 fundus images with the dimension of $2048 \times 2392$ pixels and a field-of-view of 30 degrees. Since only the A/V labels of vessel centerline are provided, we cannot train or fine-tune the model on the INSPIRE-AVR dataset. Result reported for INSPIRE-AVR in this paper is obtained by direct model inference for all of the 40 images.

During the training stage, patches with the dimension of $256\times 256$ pixels are randomly extracted from the fundus images and fed to the network. In the test stage, we extract the ordered patches at the stride of 50 and then stitch the corresponding predictions together to obtain final results. All experiments are performed on an NVIDIA Tesla P40 GPU with 24 GB of memory. The generator and discriminator are optimized alternatively with the Adam optimizer for the maximum of 30,000 iterations and a batch size of 4. The initial learning rate is set as $2\times10^{-4}$ and halved every 7,000 iterations.

\subsection{Ablation Studies}

Detailed ablation studies have been conducted to evaluate the effectiveness of different modules in the proposed framework, including the topology ranking discriminator (TR-D) and the topology preserving module with the triplet loss (TL). In addition, in order to better assess the advantage of the proposed topology ranking discriminator against general discriminators (GD) that distinguishes the real from fake masks, related experiments have also been performed. In order to compare the proposed model under the same criteria with other methods in the literature, A/V classification performance is evaluated on the segmented vessel pixels with three metrics, including accuracy (Acc), sensitivity (Sen) and specificity (Spec). By treating arteries as positives and veins as negatives, Sen and Spec reflect the model’s capability of detecting arteries and veins, respectively. Although the existing methods evaluated the A/V classification performance on the segmented vessels only, we evaluate the performance on all the ground truth artery/vein pixels, which is a relatively  stricter criterion than that on the segmented vessels, since the classification of major vessels is comparatively an easier task if capillary vessels are not segmented.

As listed in Table \ref{tab_ablation}, general discriminator improves the accuracy of A/V classification by 2.4\%, indicating the effectiveness of adversarial learning for the A/V classification task. Compared with the general discriminator, the topology-ranking discriminator exploits the ranking information in three different levels of topological connectivity maps by using ordinal classification. When the topology-ranking discriminator is adopted, the accuracy of A/V classification is improved by 3.23\%, which is higher than the general discriminator. This indicates the advantage of the proposed topology ranking discriminator over the general discriminator for the A/V classification task. Moreover, when integrating the topology preserving module with the triplet loss, the accuracy of A/V classification is increased by 3.12\% over the baseline. Finally, the proposed method with the topology ranking discriminator and the triplet loss module achieves the best performance for A/V classification with an accuracy of 95.46\%.

\begin{table}[!t]
	\centering
	\caption{The ablation study results of artery/vein classification ($\%$).}
	\label{tab_ablation}
	\begin{tabular}{p{1cm}<{\centering}p{1cm}<{\centering}p{1cm}<{\centering}|p{1.8cm}<{\centering}|p{1.8cm}<{\centering}|p{1.8cm}<{\centering}}
		\hline
		\multicolumn{3}{c|}{Combination}&\multicolumn{3}{c}{A/V Classification}\\  \hline

		GD & TR-D & TL &   Acc  & Sen & Spec  \\ \hline
		 &  & & 	91.61&	90.94 & 	92.25 \\
		\checkmark &  &  &  94.01 &	93.83 &	94.25  \\ 
	    &\checkmark  &  & 94.84	& 94.33	& 95.40   \\
	     &  & \checkmark &  94.73 &	94.30 &	95.25   \\
		 &\checkmark  &\checkmark  & \textbf{95.46} &	\textbf{94.53}	& \textbf{96.31} \\ \hline

	\end{tabular}
\end{table}

\begin{figure}[t]
	\centering
	\includegraphics[width = 12cm]{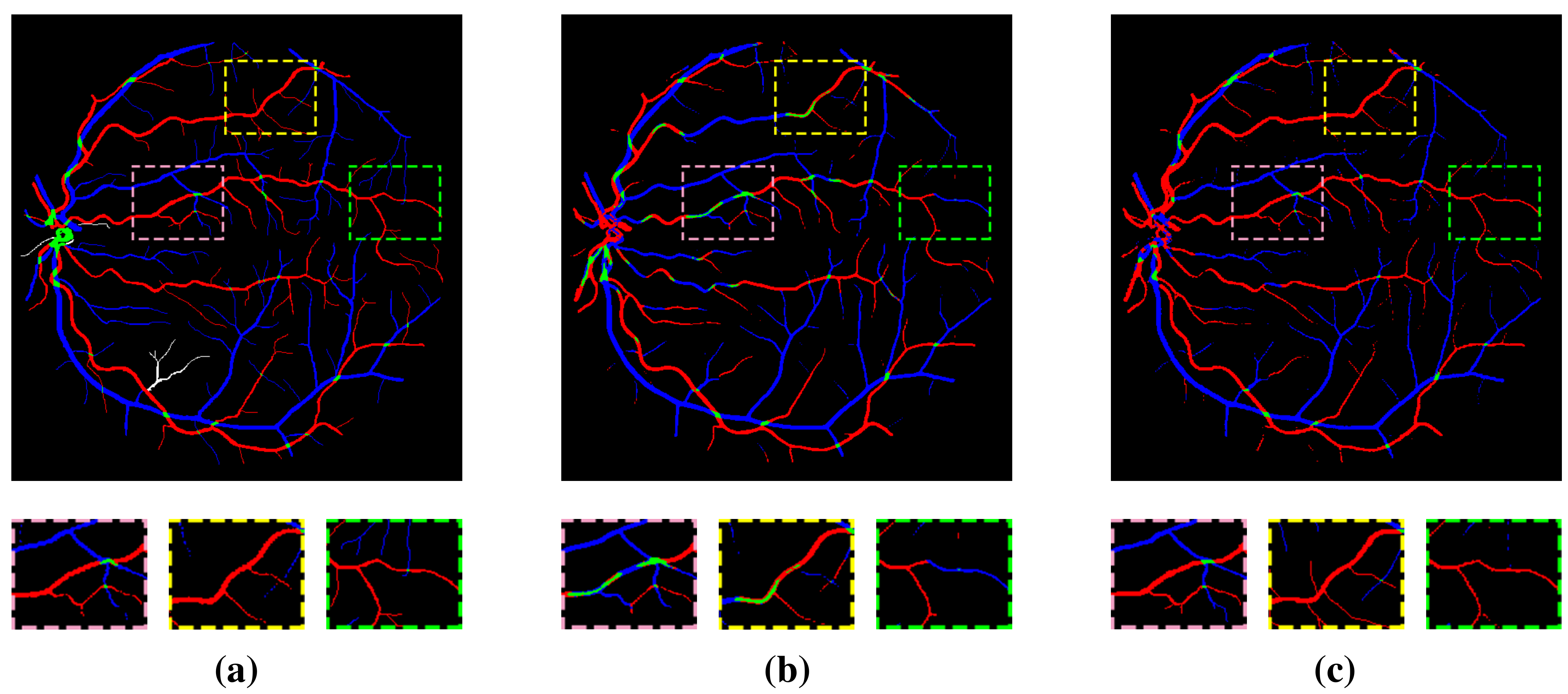}
	\caption{Comparison of model performance of the baseline and the proposed method. (a) Ground truth; (b) baseline model; (c) proposed method.} 
	\label{RepResult}
\end{figure}

Fig. \ref{RepResult} shows a representative result of the proposed model against that of the baseline. The proposed method has remarkably improved the topological connectivity of the arteries and veins, and thus achieved much better A/V classification result, as manifested in the enlarged views in the bottom row.

\subsection{Comparison with State-of-the-Art}

\begin{table}[!h]
	\centering
	\caption{Performance comparison ($\%$) of A/V classification on the AV-DRIVE and INSPIRE-AVR datasets.}\label{tab_comparison}
	\begin{tabular}{c|p{1.1cm}<{\centering}|p{1cm}<{\centering}|p{1cm}<{\centering}|p{1cm}<{\centering}|p{1cm}<{\centering}|p{1cm}<{\centering}|p{1cm}<{\centering}}
		\hline
		{} & &\multicolumn{3}{c|}{AV-DRIVE}&\multicolumn{3}{c}{INSPIRE-AVR}\\
		\cline{3-8}
		Methods&  Type & Acc& Sen  & Spec & Acc & Sen & Spec \\
		\hline
		Dashtbozorg \textit{et al.}(2014)  \cite{dashtbozorg_automatic_2014} &  & 87.4 & 90.0 & 84.0 &84.9 &91.0 &86.0 \\ 
		Estrada \textit{et al.}(2015) \cite{estrada_retinal_2015}&  & 93.5 & 93.0 & 94.1 &90.9 &91.5 &90.2 \\ 
		Zhao \textit{et al.}(2019) \cite{zhao2019retinal} & Graph & 93.5 &94.2  & 92.7 &96.4 &96.8 &95.7\\ 
		Srinidhi \textit{et al.}(2019) \cite{srinidhi2019automated} &  & 94.7 & \textbf{96.6} & 92.9 & \textbf{96.8} & \textbf{96.9} & 96.6 \\
		\cline{1-8}
		Xu \textit{et al.}(2018) \cite{xu2018simultaneous} &  & 90.0 &-  & - &79.2 &- &-\\ 
		Ma \textit{et al.}(2019)  \cite{ma2019multi} & DL & 94.5 & 93.4 & 95.5 & 91.6 & 92.4 & 91.3 \\   \cline{3-8}
		\textbf{Proposed(GT)$^*$ } & & 95.46 &	94.53	& 96.31 & - & - & - \\
		\textbf{Proposed$^\dagger$ } & & \textbf{96.29} &	95.28 &	\textbf{97.14}	& 93.40 & 89.05 & \textbf{97.29}     \\ \hline
		
		\multicolumn{8}{l}{\footnotesize{$*$: A/V classification performance is evaluated on the ground-truth vessel pixels;}} \\
		\multicolumn{8}{l}{\footnotesize{$\dagger$: A/V classification performance is evaluated on the segmented vessel pixels;}} \\
		
	\end{tabular}
\end{table}

Table \ref{tab_comparison} lists the A/V classification performance comparison between the proposed framework and related methods in the literature for both AV-DRIVE and INSPIRE-AVR datasets. Among the compared research works, the methods proposed by Xu \textit{et al.} \cite{xu2018simultaneous} and Ma \textit{et al.} \cite{ma2019multi} are deep learning based, which predict the artery and vein maps simultaneously; meanwhile the rest \cite{dashtbozorg_automatic_2014,estrada_retinal_2015,zhao2019retinal,srinidhi2019automated} are graph based methods, which extract the individual vessel tree structures first and then classify vessels into arteries and veins via a multi-step strategy. When evaluating under the same criteria with existing methods, the proposed method achieves an A/V classification accuracy of 96.29\% for the AV-DRIVE dataset, with a remarkable margin of 1.59\% over the current state-of-the-art method \cite{srinidhi2019automated}.

For the INSPIRE-AVR dataset, when comparing with deep learning based methods under the same evaluation criteria without fine-tuning, the proposed framework surpasses the methods proposed by Xu \textit{et al.} \cite{xu2018simultaneous} and Ma \textit{et al.} \cite{ma2019multi}, with an accuracy improvement of 14.2\% and 1.8\% respectively, indicating the generalization capability of the proposed model. However, when compared to the graph based methods, the performance of the proposed method is not as good as that of Zhao \textit{et al.} \cite{zhao2019retinal} and Srinidhi \textit{et al.} \cite{srinidhi2019automated}. This is primarily resulted by the fact that both of the two aforementioned methods are trained and tested on the INSPIRE-AVR dataset, meanwhile the proposed method is only trained on AV-DRIVE and then predicts on INSPIRE-AVR without further training or fine-tuning. Since the INSPIRE-AVR dataset is quite different from the AV-DRIVE dataset in terms of image color, resolution and field-of-view, it is reasonable that the performance is not as good as that on the AV-DRIVE dataset.

\section{Conclusion}

In this paper, we proposed a novel framework with topology ranking GAN and topology preserving triplet loss for the retinal artery/vein classification task. The proposed framework takes advantage of the topological connectivity of arteries and veins, to further improve the A/V classification performance. Detailed ablation and comparison studies on two publicly available datasets validated the effectiveness of the proposed method, which has achieved state-of-the-art performance on the AV-DRIVE dataset with a remarkable margin over the current best method. To the best of our knowledge, this is the first research that tries to exploit topology information for A/V classification with deep learning based frameworks. 

On the basis of the current research, future work will continue to explore the feasibility of a fully automated vascular parameter quantization system, which is expected to facilitate the clinical biomarker study of how various cardiovascular and cerebral diseases affect the retinal vessels.

\subsubsection{Acknowledgment}
This work was funded by the Key Area Research and Development Program of Guangdong Province, China (No. 2018B010111001), National Key Research and Development Project (No. 2018YFC2000702), National Natural Science Foundation of China (No. 91959108), and Science and Technology Program of Shenzhen, China (No. ZDSYS201802021814180)

\footnotesize
\bibliographystyle{splncs04}
\bibliography{paper1017}

\end{document}